\documentclass[conference]{IEEEtran}
\IEEEoverridecommandlockouts

\usepackage{cite}
\usepackage{amsmath,amssymb,amsfonts}
\usepackage{algorithmic}
\usepackage{graphicx}
\usepackage{textcomp}
\usepackage{xcolor}
\usepackage{booktabs}
\def\BibTeX{{\rm B\kern-.05em{\sc i\kern-.025em b}\kern-.08em
    T\kern-.1667em\lower.7ex\hbox{E}\kern-.125emX}}

\usepackage{amsmath,amssymb,amsfonts}
\usepackage{algorithmic}
\usepackage{graphicx}
\usepackage{textcomp}
\usepackage{makecell}
\usepackage{xcolor}
\usepackage{booktabs}
\usepackage{tikz}
\usepackage{ulem}
\usepackage{multirow}
\usepackage{pifont}
\usepackage{tcolorbox}
\usepackage{soul}
\usepackage{setspace}
\usepackage{threeparttable}
\usepackage{url}
\usepackage{subcaption}
\usepackage{graphicx}
\usepackage{hyperref}
\usepackage{diagbox}
\usepackage{enumitem}
\usepackage{listings}
\usepackage{xcolor}
\usepackage{afterpage}

\setlist[itemize]{left=0pt}

\newcommand{\finding}[2]{
    \begin{center}
    \fcolorbox{black}{gray!10}{\parbox{.97\linewidth}{
    {#2}
    }}
    \end{center}
}

\begin{document}

\title{LLMCup: Ranking-Enhanced Comment Updating with LLMs}

\makeatletter
\newcommand{\linebreakand}{%
  \end{@IEEEauthorhalign}
  \hfill\mbox{}\par
  \mbox{}\hfill\begin{@IEEEauthorhalign}
}
\makeatother

\author{
\IEEEauthorblockN{Hua Ge}
\IEEEauthorblockA{
\textit{Nanjing University} \\
Nanjing, China \\
huage@smail.nju.edu.cn}
\and
\IEEEauthorblockN{Juan Zhai}
\IEEEauthorblockA{
\textit{UMass Amherst} \\
Amherst, USA \\
juanzhai@umass.edu}
\and
\IEEEauthorblockN{Minxue Pan}
\IEEEauthorblockA{
\textit{Nanjing University} \\
Nanjing, China \\
mxp@nju.edu.cn}
\linebreakand
\centering
\IEEEauthorblockN{Fusen He}
\IEEEauthorblockA{
\textit{Nanjing University} \\
Nanjing, China \\
201250098@smail.nju.edu.cn}
\and
\IEEEauthorblockN{Ziyue Tan}
\IEEEauthorblockA{
\textit{Nanjing University} \\
Nanjing, China \\
ziyuetan@smail.nju.edu.cn}
}

\maketitle

\begin{abstract}
While comments are essential for enhancing code readability and maintainability in modern software projects, developers are often motivated to update code but not comments, leading to outdated or inconsistent documentation that hinders future understanding and maintenance. Recent approaches such as CUP and HebCup have attempted automatic comment updating using neural sequence-to-sequence models and heuristic rules, respectively. However, these methods can miss or misinterpret crucial information during comment updating, resulting in inaccurate comments, and they often struggle with complex update scenarios. 
Given these challenges, a promising direction lies in leveraging large language models (LLMs), which have shown impressive performance in software engineering tasks such as comment generation, code synthesis, and program repair. This suggests their strong potential to capture the logic behind code modifications—an ability that is crucial for the task of comment updating.
Nevertheless, selecting an appropriate prompt strategy for an LLM on each update case remains challenging. To address this, we propose a novel comment updating framework, LLMCup, which first uses multiple prompt strategies to provide diverse candidate updated comments via an LLM, and then employs a ranking model, CupRank, to select the best candidate as final updated comment. Experimental results demonstrate the effectiveness of LLMCup, with improvements over state-of-the-art baselines (CUP and HebCup) by 49.0\%–116.9\% in Accuracy, 10.8\%–20\% in BLEU-4, 4.6\% in METEOR, 0.9\%–1.9\% in F1, and 2.1\%–3.4\% in SentenceBert similarity. Furthermore, a user study shows that comments updated by LLMCup sometimes surpass human-written updates, highlighting the importance of incorporating human evaluation in comment quality assessment.

\end{abstract}


\section{Introduction}
\label{sec:introduction}

In modern software development, code comments are common and essential for readability and maintainability~\cite{9284065, DBLP:conf/iwpc/StapletonGLEWL020}. Researchers~\cite{He2019UnderstandingSC} found that over 20\% of non-blank files in 150 popular GitHub projects contain comments. Comments aid in understanding functionality, conveying design intentions, and facilitating communication among developers.

Maintaining code comments as software evolves remains a significant challenge~\cite{dagenais2014using}. Developers, often prioritizing correctness and performance, may unintentionally overlook comment updates. As software systems become more complex and undergo frequent changes, comments can quickly become outdated or inaccurate~\cite{DBLP:conf/icse/GengWD00JML24}. Such inconsistencies hinder comprehension and may increase the risk of errors during maintenance or code reviews.
To support comment consistency, various strategies and tools have been explored. Among them, CUP~\cite{DBLP:conf/iwpc/LinW0MB21} and HebCup~\cite{DBLP:conf/kbse/LiuXYL20} are notable approaches to automating comment updates. CUP employs a neural Sequence-to-Sequence model with task-specific enhancements, while HebCup builds on CUP's failure using simple heuristics to reflect code changes in comments.

Although CUP and HebCup have advanced automated comment updating, they still face notable limitations. First, their relatively low accuracy in updating comments can result in code-comment inconsistencies, potentially hindering code comprehension, maintainability, and development efficiency. 
As described in Section~\ref{sec:motivation}, both methods may produce updates that closely resemble outdated comments without reflecting code changes.
Moreover, these approaches encounter difficulties in handling complex updates that cannot be straightforwardly inferred from code edits, as demonstrated by their original analysis.
This restricts their applicability in real-world scenarios.
Another limitation is their inadequate natural language processing capability, which results in grammatically flawed reuse of words in comments or codes, or awkwardly phrased structures, ultimately reducing comment readability and developer confidence.
Finally, CUP and HebCup rely exclusively on automated evaluation metrics. Our findings suggest that the reliability of these metrics warrants further validation, especially for nuanced language tasks.
In summary, while CUP and HebCup represent valuable progress, their challenges—namely low accuracy, limited handling of complex comments, insufficient NLP capabilities, and evaluation shortcomings—highlight the need for further research to improve automated comment update systems.

A key insight is that recent advances in LLMs have shown strong capabilities in understanding and generating natural language, particularly in tasks such as code summarization, comment generation, and machine translation~\cite{Hendy2023HowGA, 10.1145/3524610.3527924, Ahmed2022FewshotTL}. These strengths present promising opportunities for enhancing comment updating by enabling more accurate and contextually appropriate modifications, thereby expected to overcome limitations of prior approaches such as Cup and HebCup.

Another important observation is the significant variability in the effectiveness of prompt strategies across different LLMs~\cite{Sun2024SourceCS} and comment update scenarios, as also evidenced by our empirical findings. This highlights the need to automatically select optimal one from the  candidate comments updated by an LLM for each comment update instance.

Motivated by these insights, we propose \textbf{LLMCup}, a comment updating framework that adopts an \textit{update-then-rank} paradigm. LLMCup first leverages an LLM to produce multiple candidate updated comments using diverse prompt strategies. To  automatically select the optimal comment, we introduce a novel ranking model, \textbf{CupRank}, which incorporates all prompt strategies, automatically ranks the corresponding candidates, and selects the top-ranked candidate as the final updated comment.

\noindent\textbf{Our main contributions are as follows:}
\begin{itemize}
    \item 
    We propose LLMCup, a novel framework for comment updating that follows an update-then-rank paradigm. It uses a base LLM with diverse prompt strategies to provide candidate updated comments respectively, followed by ranking these candidates with a ranking model, CupRank, to select the best one.
    
    \item 
    We introduce CupRank, a novel learning-based ranking model, to rank updated comments from diverse prompt strategies on the same LLM. To the best of our knowledge, this is the first work to incorporate a ranking stage into the comment updating task.
    
    \item 
    We create an augmented dataset based on the existing dataset for updated comments ranking.
    
    \item 
    We conduct extensive experiments comparing LLMCup with state-of-the-art baselines using various metrics, demonstrating its superior performance by 49.0\%–116.9\% in Accuracy, 10.8\%–20\% in BLEU-4, 4.6\% in METEOR, 0.9\%–1.9\% in F1, and 2.1\%–3.4\% in SentenceBert similarity.

    \item 
    We perform an in-depth analysis on the Accuracy of LLMCup and the state-of-the-art baselines (Cup and HebCup) across different types of comment updates. This reveals that LLMCup outperforms baselines on all types in Accuracy, and all of them are limited in handling complex update types, pointing to promising directions for future research.
    
    \item 
    We conduct a human evaluation focusing on consistency, naturalness, and helpfulness of updated comments. The results indicate that, on average, updated comments by LLMCup were rated more favorably than those by humans.
    
    \item 
    We release the LLMCup code and augmented dataset at \url{https://anonymous.4open.science/r/LLMCup}, promoting transparency and future work in comment updating.
\end{itemize}

\section{Background and Related Work}
\label{sec:relatedwork}

\subsection{Large Language Models}

LLMs have advanced remarkably in recent years as they scaled to hundreds of billions of parameters~\cite{Chowdhery2022PaLMSL} and demonstrated emergent abilities~\cite{Wei2022EmergentAO} on completing multiple complex task, such as Question Answering, Reasoning, Math/Science and Coding~\cite{Jiang2023Mistral7}.
Due to their strong capability on in-context learning, LLMs are widely applied in the filed of programming-related tasks, including code generation~\cite{Liu2023IsYC, Chen2021EvaluatingLL, Deng2022RecentAI, 10.1145/3533767.3534390, Mu2023ClarifyGPTEL}, software testing~\cite{Deng2023LargeLM, Xia2023Fuzz4ALLUF}, program verification~\cite{Si2018LearningLI, Janssen2023CanCS}, code summarization~\cite{Ahmed2022FewshotTL, Sun2024SourceCS} and comment generation~\cite{DBLP:conf/icse/GengWD00JML24, 10.1145/3524610.3527924}. 
There have been a long list of language models rising in recent years.
In this paper, we focus on some prevalent models, such as GPT-4o~\cite{Hendy2023HowGA, web:GPT-4o}, Llama (Llama3:8b, CodeLlama:7b)~\cite{web:Llama3, Rozire2023CodeLO}, DeepSeek-Coder-v2:16b~\cite{zhu2024deepseek}, Gemma:7b~\cite{web:Gemma}, Mistral:7b~\cite{Jiang2023Mistral7}. 
Compared to these work, our goal is to employ LLMs for automated comment update, which is crucial to software maintenance. 

\subsection{Comment Updating}
Changing codes without updating comments might brings potential bugs into the program~\cite{10.1145/1137983.1138030}, which makes it important to keep comment consistent with the latest codes. 
The expectation to reduce  potential bugs and developers' workload motivate a large amount of work on automatic comment update, which can categorized into two types: heuristic-based and machine learning-based. 
For heuristic-based approaches, Bo et al.~\cite{DBLP:conf/iwpc/LinW0MB21} designed heuristic rules from code comment update cases and implemented HebCup which can directly update comments using heuristics.  
As for machine learning based approaches, Liu et al.~\cite{DBLP:conf/kbse/LiuXYL20} introduced CUP which leverage neutral sequence-to-sequence model to learn and update the comments. 
Zhu et al.~\cite{Zhu2022HatCUPHA} proposed HatCUP based on RNN-based encoder-decoder architecture, which generate a sequence of comment edit actions.
Guo et al.~\cite{Guo2022DeepJC} uesed self-attention and positional encoding mechanism to capture long-term and non-temporal dependencies in the source codes, which help models update the comments based on old comments and code editing.
Compared to these work, we try enhancing comment update performance with LLMs and prompt engineering. 

\subsection{Comment Generation}
\label{sec:related work - comment Gen}
Generating new comments for updated codes from scratch is also an important way to eliminate inconsistencies between comments and codes. 
In automatic comment generation, a large part of research focus on applying predefined templates to generate comments~\cite{Haiduc2010OnTU, Eddy2013EvaluatingSC, Haiduc2010SupportingPC, Sridhara2010TowardsAG}, while many others are IR-based~\cite{Wong2015CloComME, Chatterjee2017ExtractingCS, Jiang2017AutomaticallyGC, Wong2013AutoCommentMQ}, e.g. Wong et al.~\cite{Wong2015CloComME} prooposed an approach of exacting existing corresponding descriptions and codes from Question and Answering sites and leverage them for similar code segments. 
There also have been great efforts generating comments based on neutral network\cite{Zheng2018CodeAttentionTS, Iyer2016SummarizingSC, Wang2017AutomaticallyGN, Vedantam2014CIDErCI, Wan2018ImprovingAS, Cho2014OnTP, Roehm2012HowDP, Allamanis2016ACA}. 
For example, Zheng et al.~\cite{Zheng2018CodeAttentionTS} took advantage of code constructs such as critical statements, symbols and keywords, to help neural network translate source codes into comments. 
Wan et al.~\cite{Wan2018ImprovingAS} incorporate an abstract syntax tree structure and sequential content into a deep reinforcement learning framework for comment generation.
These machine learning-based approaches have demonstrated significant performance improvement on comment generation task.
Therefore, researchers also explored LLMs' capabilities on generating comments~\cite{DBLP:conf/icse/GengWD00JML24, Sun2024SourceCS, 10.1145/3524610.3527924}. 
Geng et al.~\cite{DBLP:conf/icse/GengWD00JML24} utilized LLMs to generate comments which reveal developers' intents. 
Different from these work, our approach is to evaluate LLMs' abilities to update the existing out-dated comments of codes. 

\subsection{Inconsistent Comment Detection}
Many researchers focus on how to detect the code-comment inconsistencies to reduce the potential harmfulness. 
For many previous work, they tend to use program analysis techniques and target for specific type of comments~\cite{Tan2007icommentBO, Khamis2010AutomaticQA, Tan2012tCommentTJ}. 
For example, Tan et al.~\cite{Tan2012tCommentTJ} tested the comments and methods related to null value and exceptions in Java projects, by employing static analysis and automated testing. 
In addition, researchers also took efforts on utilizing machine learning techniques for analyzing inconsistency. 
Fraco~\cite{Ratol2017DetectingFC} identify  outdated or incorrect comments, using machine learning algorithms, specifically decision trees and support vector machines (SVM), to analyze historical code changes and comment updates. 
Compared to these work, our expectation is to help developers automatically update the inconsistent comments when they detect any such comments.

\section{Motivation}
\label{sec:motivation}

\begin{figure}[]
     \centering
     \vspace{-15pt}
    \includegraphics[width=0.5\textwidth]{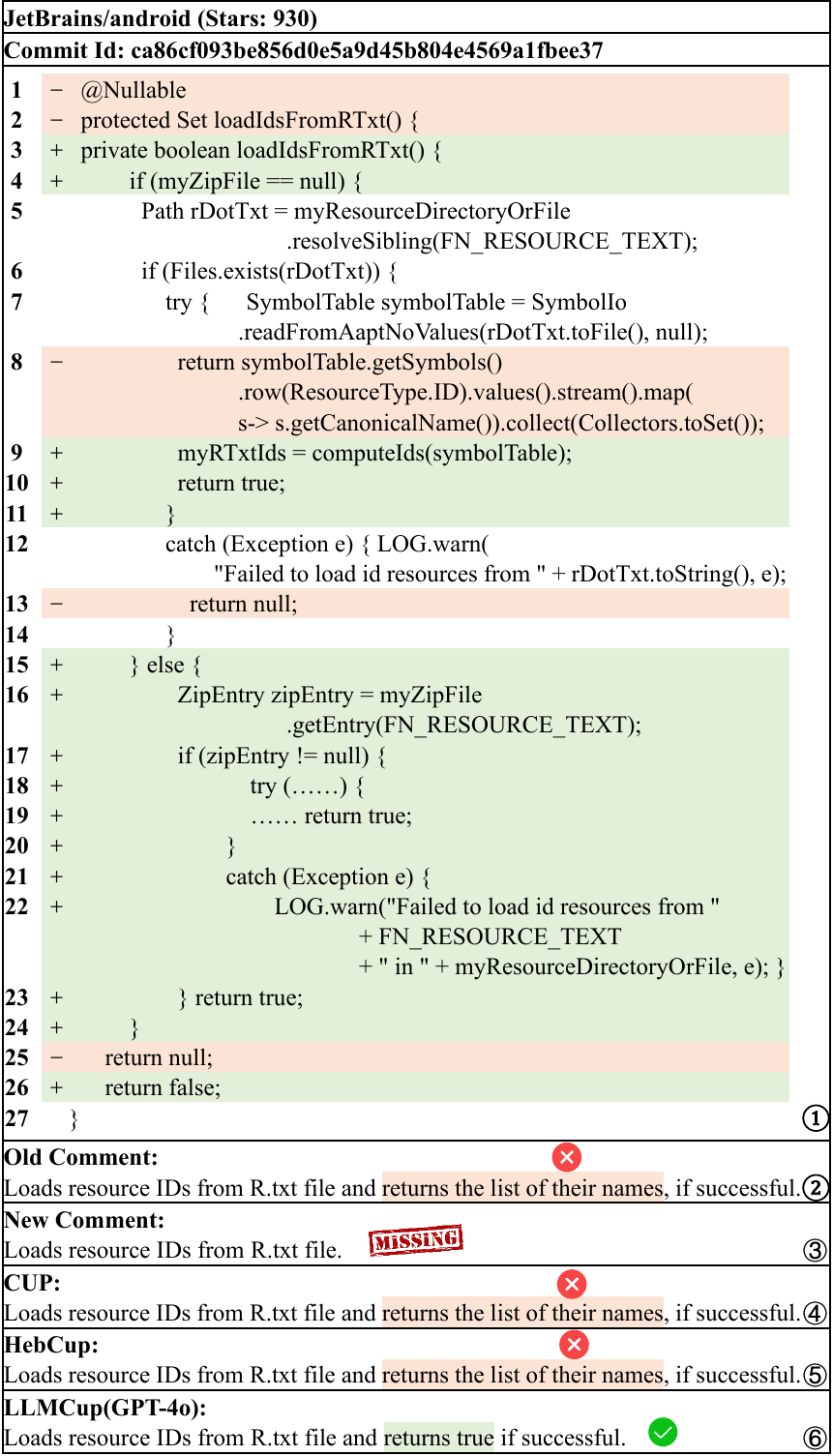}
     \caption{Motivating Example}     
    \vspace{-10pt}
    \label{fig:v1_motivation_example}
     \vspace{-10pt}
\end{figure}

This section describes the motivation of  LLMCup, which follows an update-then-rank paradigm, through answering two questions as follow:

\textbf{Why Use LLMs for Comment Updating?}
Fig.~\ref{fig:v1_motivation_example} shows an example we found in a real-world GitHub repository of JetBrains/android\cite{web:JetBrains}. 
As \textcircled{1} presents, the code changed and the current logic of the method is to return a boolean value (at line 3,10,23,26). 
The original comment in \textcircled{2} states that the method loads resource IDs and returns their names, but this no longer reflects the updated implementation. The new comment in \textcircled{3}, written by developers, omits the change in return value type. Such omissions can lead to misuse, longer debugging time, and decreased code readability and maintainability.

HebCup and CUP are two state-of-the-art comment update approaches. \textcircled{4} and \textcircled{5} are the comments updated by them, respectively. However, both incorrectly update the comment as ``\texttt{returns the list of their names}'', which is outdated and misses the change in return type. In contrast, the updated code returns a boolean under varying conditions. Such erroneous updates or direct copies introduce code-comment inconsistencies that can mislead developers, causing false assumptions, longer debugging, and potential maintenance errors, ultimately undermining code reliability and maintainability.

LLMs have shown strong performance in various SE tasks, including comment generation~\cite{wu2024xinyu}, comment-code inconsistency detection~\cite{zhang2024detecting}, and software testing~\cite{chen2024chatunitest}. Building on this, we explore their potential for comment updating by prompting LLMs with a carefully designed template (Fig.~\ref{fig:prompt}) to update outdated comments.
For example, 
The comment in \textcircled{6}, updated by LLMCup with GPT-4o,  accurately summarizes the updated code's core functionality and specifies the return behavior. Notably, this return detail is absent in the developer-updated comment and is mishandled by existing approaches. 

Motivated by above observations, we employs LLMs to update comments.

\textbf{Why Rank?}
As reviewed in~\cite{Sun2024SourceCS} (Section~\ref{sec:related work - comment Gen}), no single prompt strategy consistently outperforms others across all LLMs, aligning with our observations in comment updating. 
Moreover, no strategy is universally superior across all update cases. 
For instance, GPT-4o (0-shot) may outperform GPT-4o (5-shot) in some cases, while the reverse holds in others.
Furthermore, effective ranking could reduce the randomness in LLMs' outputs during comment updating, by prioritizing higher-quality candidate updated comments.

These findings motivate the addition of a ranking stage after the LLM-based updater in LLMCup, enabling automatic integration of updated comments from an LLM without  Compromising performance.

\section{Methodology}
\label{sec:methodology}

\subsection{Overview}

\begin{figure*}[]
     \centering
     \vspace{-20pt}
    \includegraphics[width=0.7\textwidth]
    {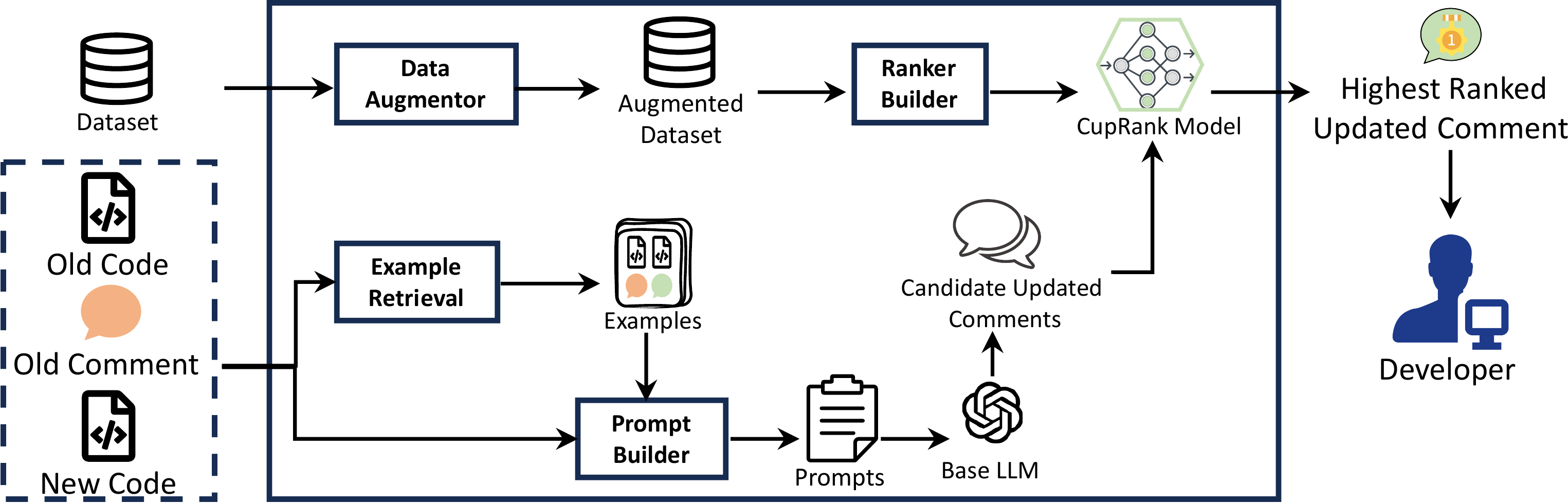}
     \setlength{\abovecaptionskip}{0.cm}
     \caption{Overview of Framework
     }   
     \label{fig:Framework}
     \vspace{-10pt}
\end{figure*}

\begin{figure}[]
     \centering
     \vspace{-10pt}
    \includegraphics[width=0.4\textwidth]{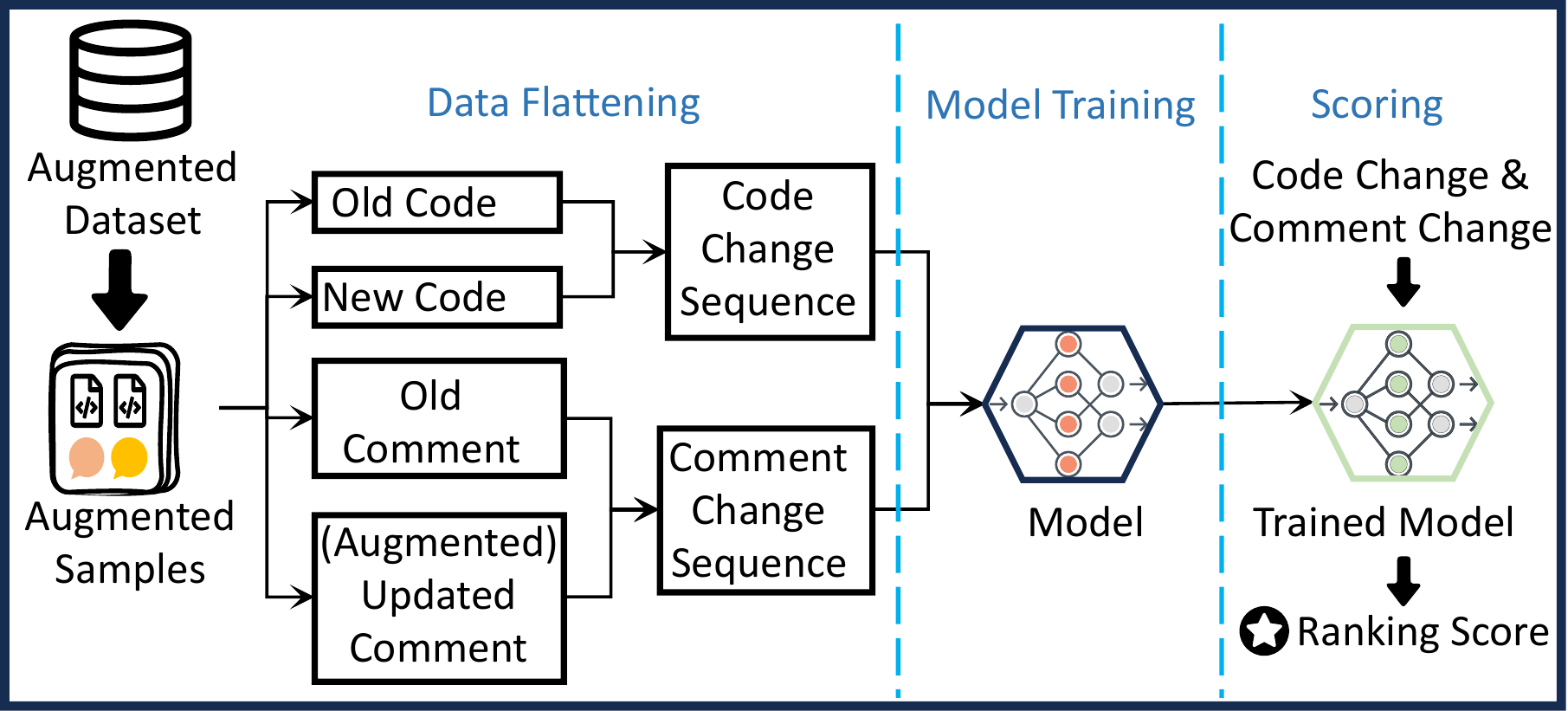}
     \setlength{\abovecaptionskip}{-0cm}
     \caption{Training of CupRank}   
     \label{fig:Training}
     \vspace{-20pt}
\end{figure}

As illustrated in Fig.~\ref{fig:Framework}, \textsc{LLMCup} is a framework for updating comments by LLMs and ranking. It follows an \textit{update-then-rank} paradigm: multiple candidate comments are produced, respectively, via diverse prompting strategies with a same LLM, and the best one is selected by a learned ranker.

The first input in Fig.~\ref{fig:Framework} is the existing comment updating dataset. It is used for \textit{Data Augmentation}, where LLMs (in section~\ref{Experimental LLMs}, excluding GPT-4o) under diverse prompt strategies are employed by the \textit{Data Augmentor} to construct augmented dataset from the original.
In particular, augmented dataset consists of augmented sample groups, each of which contains a positive sample (a sample from the original dataset) and more than one negative samples (derived from this positive sample by the \textit{Data Augmentor}).
An  positive (original) sample consists of an old code, an old comment, a new code and a new (namely updated) comment, while the corresponding negative sample consists of the same old code, old comment, new code, but an augmented updated comment, which is not identical to the new comment in this original sample and updated by one of the \textit{Data Augmentor}'s LLMs under one of prompt strategies (e.g., 0-shot).
However, not every original sample can yield negative samples, because all the updated comments produces by the \textit{Data Augmentor} for a given original sample may be identical to its new comment.  
As a result, the augmented dataset is constructed, and the original samples with no corresponding negative samples are discarded.

Notably, we do not filter semantically similar negatives, as ranking models learn relative order, not strict dissimilarity~\cite{RanknetBurges2005}. Such negatives often act as hard examples that improve representation learning~\cite{robinson2021contrastivelearninghardnegative, xiong2020approximatenearestneighbornegative}. Moreover, identifying semantic equivalence is costly and ambiguous, and avoiding it improves scalability.

Based on the augmented dataset, the \textit{Ranker Builder} trains \textit{CupRank} using a listwise loss over each sample group. As shown in Fig.~\ref{fig:Training}, each sample is flattened into code and comment change sequences. The model learns to output ranking scores reflecting alignment between code/comment changes.

The second input is a real sample (old/new code and old comment) to infer. \textit{Example Retrieval} finds similar examples based on new code similarity. \textit{Prompt Builder} creates multiple prompts using these examples and a prompt template under various strategies (e.g., 0-shot, 5-shot). The base LLM (e.g., GPT-4o) produces candidate updated comments, respectively, which are then ranked by \textit{CupRank}. The top-ranked one is returned to the developer.

\begin{figure*}[]
     \centering
     \vspace{-10pt}
     \includegraphics[width=0.7\textwidth]{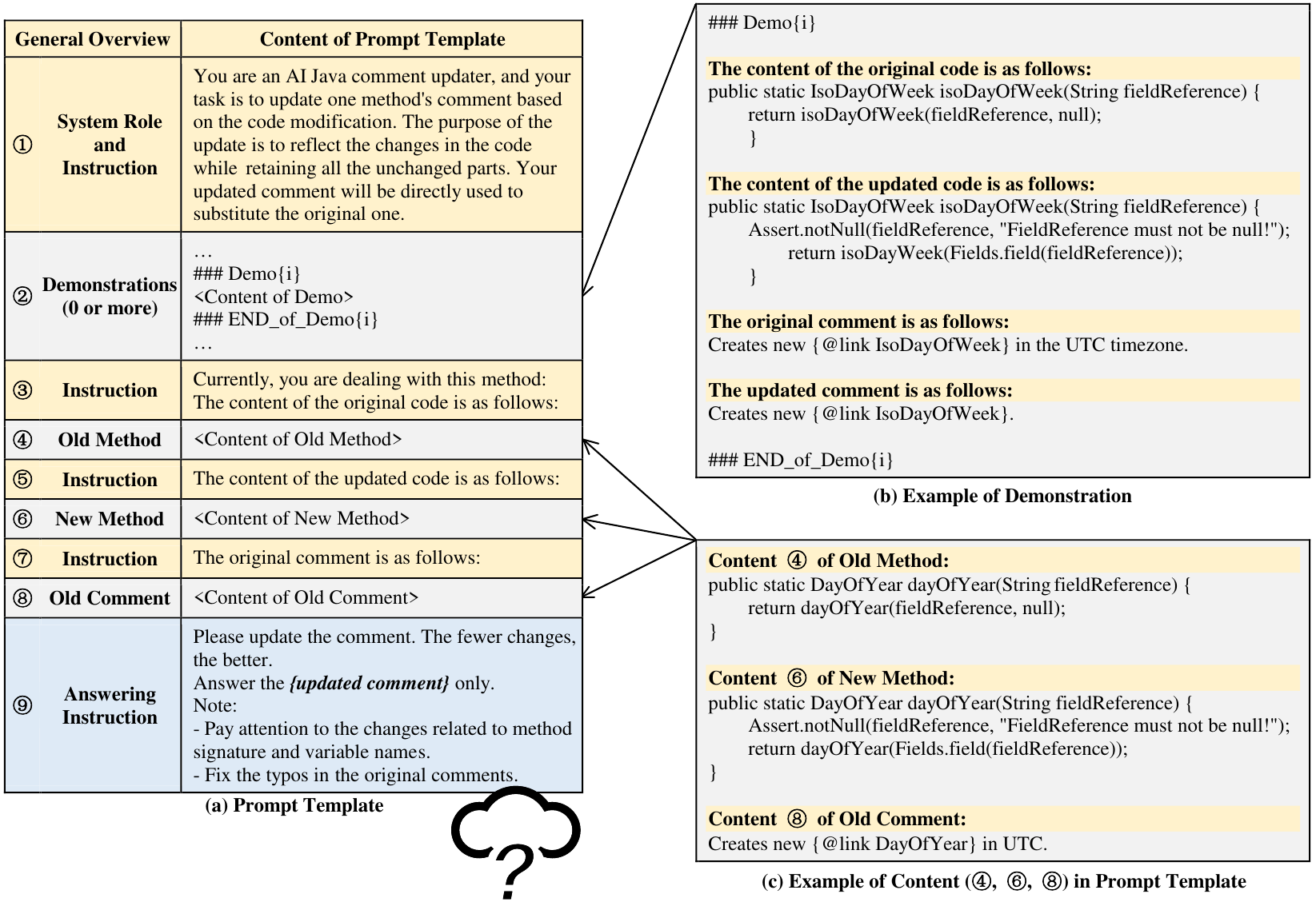}
     \setlength{\abovecaptionskip}{0.cm}
     \caption{Structure and Example of the Prompt Template for Comment Updating}
     \label{fig:prompt}
     \vspace{-10pt}
\end{figure*}

\subsection{Prompt Template}
\label{sec:prompt template}

To enable LLMs to automatic update comment, we design a structured prompt (Fig.~\ref{fig:prompt}) with the following components:

\begin{itemize}
    \item \textcircled{1} \textbf{System Role \& Instruction (fixed)}: Defines the task of updating outdated comments.
    \item \textcircled{2} \textbf{Demonstrations (optional and instance-specific)}: Retrieved from the training corpus via semantic similarity between new codes of instances using a pretrained encoder~\cite{reimers2019sentence}. Prior works~\cite{min2022rethinking, Retrieval-Based10172590, rubin-etal-2022-learning, nashid2023retrieval} show these guide LLMs better than random ones.
    
    \item \textcircled{3} \textbf{Old Method Instruction (fixed)} \& \textcircled{4} \textbf{Old Method (instance-specific)}: Introduce and present the outdated code.
    
    \item \textcircled{5} \textbf{New Method Instruction (fixed)} \& \textcircled{6} \textbf{New Method (nstance-specific)}: Introduce and present the revised code.
    
    \item \textcircled{7} \textbf{Old Comment Instruction (fixed)} \& \textcircled{8} \textbf{Old Comment (instance-specific)}: Display the original comment.
    
    \item \textcircled{9} \textbf{Answering Instruction (fixed)}: Specifies the response format, enforcing two rules that empirically improve LLM outputs:
    (1) \texttt{The fewer changes, the better.} Promotes minimal edits for clarity and alignment with software maintenance practices.
    (2) \texttt{Fix typos in the original comments.} Improves readability and precision; LLMs are effective at such corrections.
\end{itemize}

\subsection{Data Augmentation}
\label{data aug}

\textit{CupRank} learns to rank by explicitly modeling the relative order between positive and negative examples in augmented training samples.
However, the existing dataset only contains positive samples (i.e., old/new code, old comment, and ground truth comment), lacking such contrastive examples.

To address this, we use LLMs with diverse prompting strategies to generate a large-scale augmented dataset. As shown in Fig.~\ref{fig:DataAug}, given a sample (step 1) with old/new code and comment, \textsc{LLMCup} removes the new comment to form the inference input. In step 2, it retrieves $k$ similar examples based on new code similarity to construct prompts using various templates. Step 3 uses LLMs (see Sections~\ref{sec:Experiment Settings} and~\ref{Experimental LLMs}) to generate candidate updated comments. These, along with the ground truth, form the augmented samples after deduplication.

In step 4, \textsc{LLMCup} repeats steps 2--3 across the dataset, producing augmented samples with old code, old comment, new code, and an updated comment. A sample is labeled \textit{positive} if its updated comment matches the ground truth; otherwise, \textit{negative}. This yields an augmented dataset combining LLM-generated and original data.

\begin{figure}[t]
     \centering
     \vspace{-5pt}
    \includegraphics[width=0.4\textwidth]{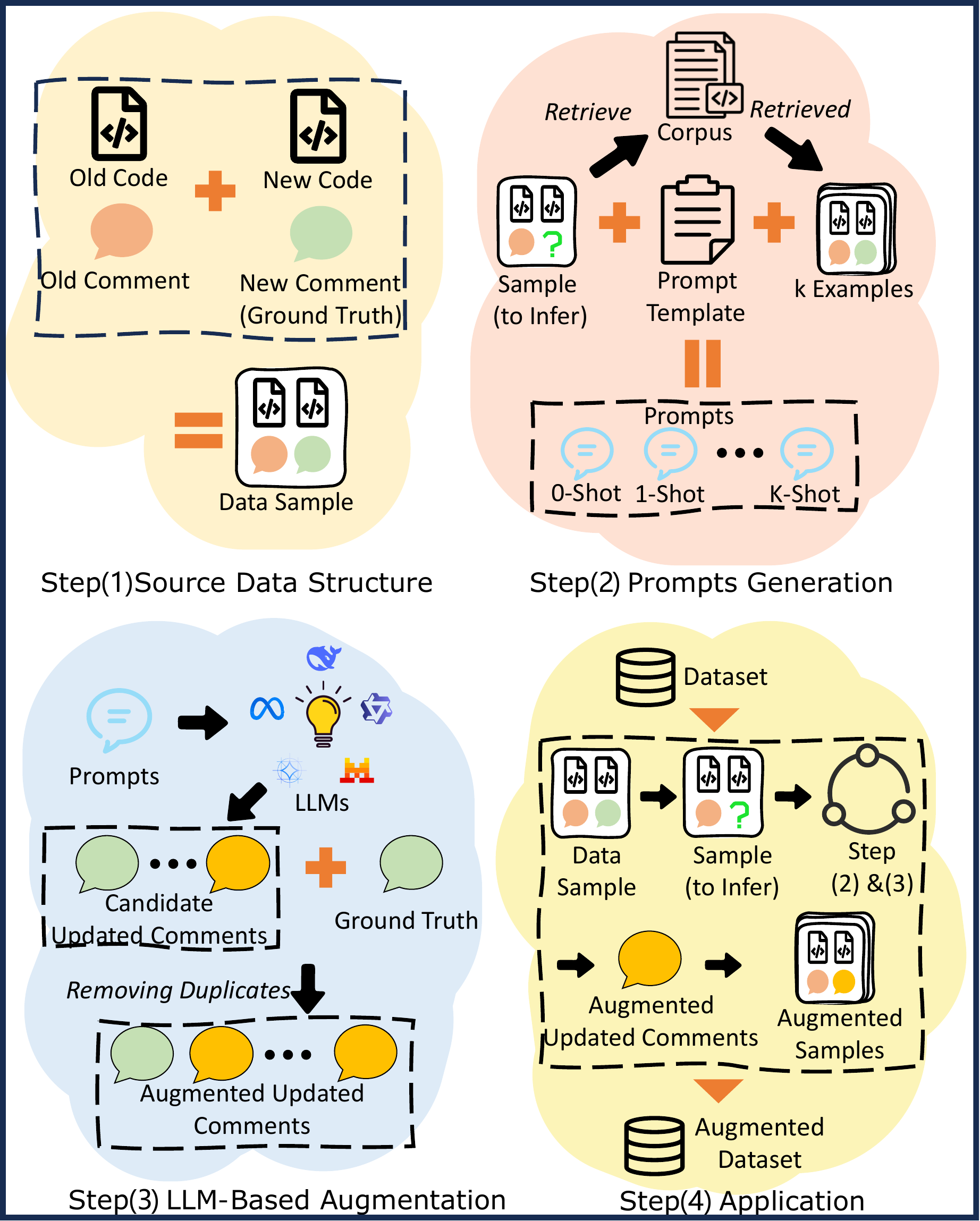}
     \setlength{\abovecaptionskip}{0.cm}
     \caption{Data Augmentation}   
     \label{fig:DataAug}
     \vspace{-15pt}
\end{figure}

\subsection{Data Flattening}
\label{flattening}

\begin{figure*}[]
     \centering
     \vspace{-10pt}
    \includegraphics[width=0.8\textwidth]{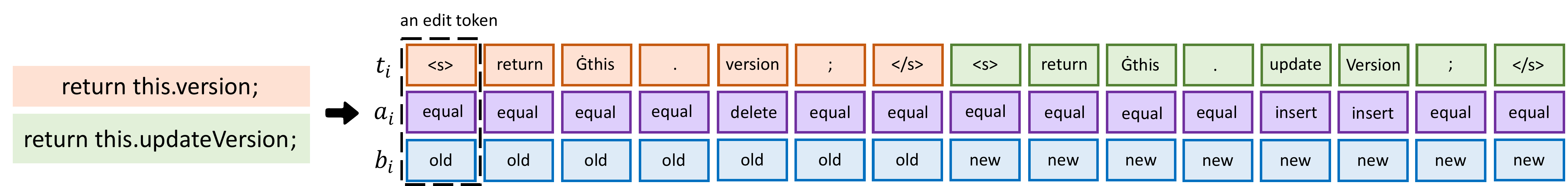}
     \setlength{\abovecaptionskip}{0.cm}
     \caption{Convert Code/Comment Change to Change Sequence in the stage of Data Flattening}   
     \label{fig:change2seq}
     \vspace{-15pt}
\end{figure*}

To represent comment update samples effectively, we address three challenges: (1) heterogeneous components (code vs. comments), (2) multiple components (old/new code and comments), and (3) implicit modification info (code and comment changes).
To tackle these, we propose \textit{Data Flattening} during \textit{CupRank} training (Fig.~\ref{fig:Training}), which converts each sample into two sequences of edit tokens (Fig.~\ref{fig:change2seq})

First, for heterogeneous inputs, \textit{Data Flattening} uses CodeBERT~\cite{Feng2020CodeBERT} for unified tokenization and embedding of both code and comments. As illustrated in Fig.~\ref{fig:change2seq}, CodeBERT splits text into subword units (e.g., ``updateVersion'' $\rightarrow$ ``update'', ``Version''), and inserts special tokens like ``\textless{}s\textgreater{}'', ``\textless{}/s\textgreater{}'', and ``Ġ'', meaning start, end and prefixed space.

Second, to manage multiple components, we concatenate tokens of old/new code and comments respectively, resulting in two flattened sequences for code/comment change per sample.

Third, to represent implicit modifications, a token-level diff~\cite{python-difflib} identifies edit operations and token origins.

Formally, for each token $t_i$:
the edit operation is defined as $a_i \in \{\texttt{equal}, \texttt{insert}, \texttt{delete}\}$;
the token origin is defined as $b_i \in \{old, new\}$.
Thus, an edit token is a triplet as follow: 
\begin{equation}
    \text{edit token}_i = (t_i, a_i, b_i)
    \label{eq:edit_token}
\end{equation}
The embedding of $\text{edit token}_i$ is defined as:
\begin{equation}
    concatenate(e_{t_i}, e_{a_i}, e_{b_i})
    \label{eq:edit_token_emb}
\end{equation}
where, 
$e_{t_i}$ denotes the embedding of $t_i$ obtained from CodeBERT; $e_{a_i}$ is the one-hot encoding of $a_i$, specifically [1,0,0] for \texttt{equal}, [0,1,0] for \texttt{insert}, and [0,0,1] for \texttt{delete}; $e_{b_i}$ is the origin flag (0 for old, 1 for new). The function $concatenate()$ is used to combine $e_{t_i}$, $e_{a_i}$, and $e_{b_i}$ into a single embedding for $\text{edit token}_i$.

\subsection{CupRank Model}
\label{subsec:cuprank}

\begin{figure}[]
     \centering
     \vspace{-10pt}
    \includegraphics[width=0.49\textwidth]{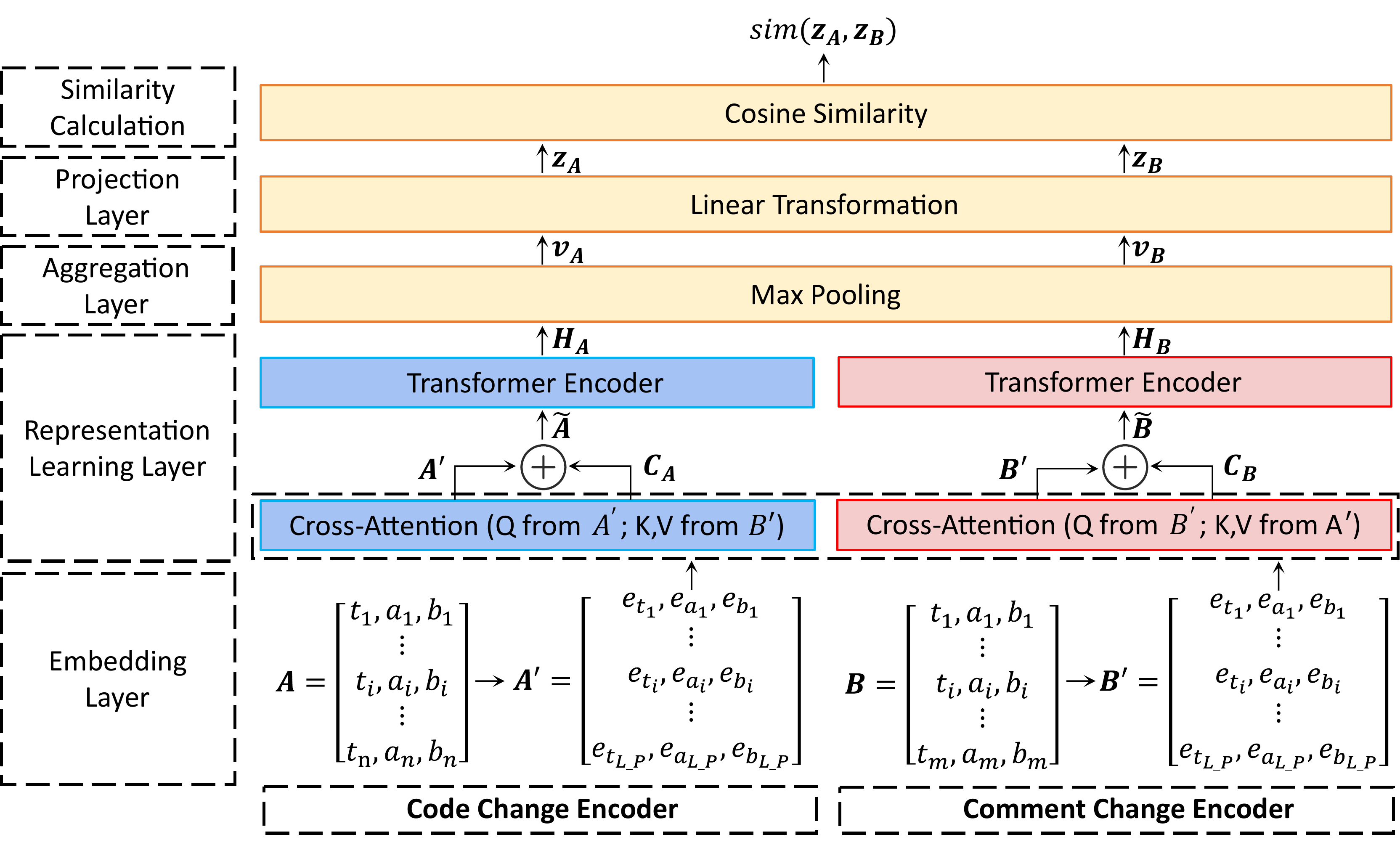}
     \setlength{\abovecaptionskip}{-0.3cm}
     \caption{Architecture of CupRank Model}   
     \label{fig:Architecture}
     \vspace{-15pt}
\end{figure}

To model the relative
order between positive and negative samples with the same old/new code and old comment,
we propose \textbf{CupRank}, a neural ranking model trained on the augmented dataset.
As shown in Figure~\ref{fig:Architecture}, CupRank uses a dual-encoder (code and comment change encoders) to process the flattened input sequences $A$ and $B$.
Each encoder consists of four layers: Embedding, Representation Learning, Aggregation, and Projection.

The Embedding Layer transforms edit token sequences $A$ and $B$ into embeddings $A'$ and $B'$ (Equations~\ref{eq:edit_token},~\ref{eq:edit_token_emb}).

Representation Learning Layer, use two cross-attention to extract two-way relationships between code/comment changes:
\begin{equation}
\label{cross1}
C_A = \text{Attention}(Q=A', K=B', V=B')
\end{equation}
\begin{equation}
\label{cross2}
C_B = \text{Attention}(Q=B', K=A', V=A')
\end{equation}
followed by residual addition:
\begin{equation}
\label{cross_attens}
\tilde{A} = A' + C_A, \quad \tilde{B} = B' + C_B.
\end{equation}

Then, a 2-layer Transformer Encoder refines the contextual token-level representations, converting $\tilde{A}$, $\tilde{B}$ to $H_A$, $H_B$.
Max-pooling compresses outputs to fixed-size vectors: from $H_A$, $H_B$ to $v_A$, $v_B$.
A linear Projection Layer maps them to a shared space: from $v_A$, $v_B$ to $z_A$, $z_B$.
Cosine similarity defines the ranking score:
\begin{equation}
\text{sim}(z_A, z_B) = \frac{z_A \cdot z_B}{\|z_A\| \|z_B\|}.
\label{eq:sim}
\end{equation}
Loss is computed on a positive sample \(X_0\) and \(N\) negatives:
\begin{equation}
s_i = \exp\left(\text{sim}(z_A(X_i), z_B(X_i)) / \lambda \right)
\label{eq:logit}
\end{equation}
\begin{equation}
\mathcal{L} = -\log\left( \textstyle s_0 / (s_0 + \sum_{i=1}^{N} s_i) \right).
\label{eq:loss}
\end{equation}

Here, \( \lambda\) setted to 0.07, to controls similarity distribution smoothness. The objective encourages CupRank to rank the positive samples higher than corresponding negative ones.

\subsection{Inference Pipeline}

Given a sample to infer,
the \textit{Inference Pipeline} is:

(1) \textbf{LLM-Based Comment Updating}:  
For the sample (to infer), candidate comments are updated, respectively, by the base LLM with prompts of diverse strategies(e.g. 0-, 1-, 3-, 5-shot).
(2) \textbf{Candidates Ranking}:  
   The candidate updated comments with the sample (to infer) are then passed to the CupRank model, 
   which computes ranking scores. 
   Based on these scores, the candidates are ranked accordingly.
(3) \textbf{Best Candidate Selection}:  
   The candidate with the highest ranking score is selected as the final output. 

Overall, 
LLMCup automatically select optimal one from the  candidate comments updated by an LLM for each comment update instance via CupRank.

\section{Study Design}
\label{sec:expsetup}

\subsection{Research Questions}

We pose the following research questions to evaluate the effectiveness of LLMCup.

\textbf{RQ1: How does LLMCup perform compared to state-of-the-art comment updating baselines?}  
We assess LLMCup's effectiveness by compare it to leading approaches.

\textbf{RQ2: Does the LLMCup improve over using diverse prompt strategies alone?}  
We evaluate CupRank's value by comparing LLMCup to direct LLM-based comment updates.

\textbf{RQ3: How do ranking strategies influence LLMCup’s performance, and what drives CupRank's effectiveness?}  
We systematically compare CupRank with alternatives and conduct ablations to identify key performance contributors.

\textbf{RQ4: Under what conditions does LLMCup excel or struggle?}  
We analyze scenarios of strong and weak performance to understand LLMCup's strengths and limitations.

\textbf{RQ5: How does the temperature setting affect LLM performance in comment updating?}  
We examine how varying temperature values influence update quality, identifying optimal settings for this task.

\textbf{RQ6: To what extent do humans prefer LLM-updated comments over ground truth?}  
We conduct a human study to assess whether LLM outputs surpass human comments.

\subsection{Baselines}
\label{baselines}

LLMCup is evaluated against two baselines: Comment Update for update quality, and Ranking for \textit{CupRank}.

\noindent
\textbf{\textit{Comment Updating Baselines}}

\noindent
\textbf{CUP}~\cite{DBLP:conf/kbse/LiuXYL20}: 
A neural seq2seq model for comment updating. It learns comment evolution patterns from historical code-comment co-changes. It incorporates a tailored tokenizer and co-attention mechanism to handle characteristics of the task.

\noindent
\textbf{HebCUP}~\cite{DBLP:conf/iwpc/LinW0MB21}: 
A simple yet effective heuristic alternative to CUP that updates comments via token-level edits.

\noindent
\textbf{\textit{Ranking Baselines}}

\noindent
\textbf{Random}: Randomly selects a candidate comment from the set, serving as a lower-bound reference for ranking performance.

\noindent 
\textbf{Self-Ranking}: An LLM-based ranker, inspired by recent study~\cite{LLM4RankSun2023InstructionDM, LLM4RankSun2023IsCG}. 
It prompts an LLM to rank comments updated by itself. 
The \textit{Self-Ranking Prompt Template} is as follow:

\begin{lstlisting}[basicstyle=\ttfamily\scriptsize, xleftmargin=0pt]
###You are a Comment Judge - an expert software engineer tasked with evaluating and ranking candidate updated comments.  ###Currently, you are dealing with this method: ###The content of the original code is as follows: ###{old_method}The content of the updated code is as follows:{new_method} ####The original comment is as follows: ###{old_comment} ###Please evaluate the following candidate updated comments provided by different experts: ###{"Expert 1": {"comment": {candidate 1} }, ..., "Expert k": {"comment": {candidate k}}} ###Please rank candidates from best (top-1) to worst (top-{k}) and return the names of corresponding experts in the following JSON format, with no additional explanation: ###{"top-1": "here_is_Expert_Name", ....,}
\end{lstlisting}
where ``\texttt{\{\}}'' means placeholder and ``\texttt{\#\#\#}'' means line break.

\noindent
\textbf{RankNet}~\cite{RanknetBurges2005}: A widely-used  ranking baseline~\cite{RanknetTask1Song2014, RanknetTask2_9054702}. 
We design a task-specific adaptation of RankNet (originally a simple loss with two fully-connected layer) to handle comment updating. 
Inputs are token embeddings (from CodeBERT~\cite{Feng2020CodeBERT}) of old/new code, old comment, and a updated comment, concatenated and fed into a feedforward network with two fully-connected layers, max-pooling, and a sigmoid for ranking.

\subsection{Dataset}
\label{Dataset}

We use a refined version of the CUP dataset~\cite{DBLP:conf/kbse/LiuXYL20}, originally built from 1,500 Java repositories on GitHub~\cite{wen2019large} to study code-comment inconsistencies. CUP splits data into training (80\%), validation (10\%), and testing (10\%) sets, removing duplicates across splits. HebCup further filters trivial revisions (e.g., case changes, lexical variations, typos). We additionally remove test samples duplicated in the validation set to prevent leakage. The final dataset contains 80,591 training, 8,827 validation, and 8,125 testing samples.

\subsection{Experimental LLMs}
\label{Experimental LLMs}

To evaluate LLM performance in comment updating and \textsc{LLMCup}'s generalizability, we selected 7 models, one commercial and six open-source—based, based on code-related relevance and Hugging Face popularity. The models are:

(1) \textbf{GPT-4o}\cite{web:GPT-4o}: OpenAI's commercial model, comparable to GPT-4 Turbo in code and text.
(2) \textbf{Code Llama-7B}\cite{Rozire2023CodeLO}: Meta's general-purpose code synthesis model.
(3) \textbf{DeepSeek-Coder-v2-16B}\cite{zhu2024deepseek}: Strong open-source coding model, competitive with GPT-4 Turbo.
(4) \textbf{Qwen2.5-Coder-7B}\cite{hui2024qwen25codertechnicalreport}: Fine-tuned for code generation and comprehension.
(5) \textbf{Mistral-7B}\cite{Jiang2023Mistral7}: Known for reasoning and code generation.
(6) \textbf{Llama3-8B}\cite{web:Llama3}: Performs well on both language and code tasks.
(7)\textbf{Gemma-7B}\cite{web:Gemma}: Lightweight model effective in tasks of summarization, question answer, and code generation.

\subsection{Experiment Settings}
\label{sec:Experiment Settings}

We standardize configurations across evaluations for reproducibility and fair comparison:
(1) \textbf{Prompting Strategies.}
We use $k$-shot prompting with $k \in \{0,1,3,5\}$ to balance context richness and input length. 
0-/1-shot serve as minimal baselines; 3-/5-shot assess scaling with more context. 
Prompts contain up to five examples due to LLMs' input window size limits. 
All LLMs are set to temperature 0.2 (based on results in Section~\ref{sec:QA5}).
Training dataset are used as corpus for example retrieval. 
(2) \textbf{Resource Limitations.}
Due to budget constraints,
RQ5 uses 1,000 randomly sampled test cases, while GPT-4o is excluded from the \textit{Data Augmentation} stage.
(3) \textbf{Model Configuration.}
\texttt{CupRank} is learning rate of 1e-4 and batch size of 8. 
Checkpoints are saved every 5,000 instances; the final one is selected by lowest validation loss. 
Training uses fixed seeds for reproducibility.
We encode inputs of CupRank and RankNet with CodeBERT (\texttt{codebert-base})~\cite{Feng2020CodeBERT}.

\subsection{Evaluation Metrics}
\label{sec:Evaluation Metrics}

To evaluate comment updating,  
we report per-metric averages following prior work~\cite{DBLP:conf/kbse/LiuXYL20, DBLP:conf/iwpc/LinW0MB21}  
and related tasks~\cite{haiduc2010use, DBLP:conf/iwpc/HuLXLJ18, DBLP:conf/acl/IyerKCZ16, Wan2018ImprovingAS}.  
Let $C_{up}$ be the updated comment and $C_{gt}$ the ground truth.  
Metrics are detailed below.

\noindent
\begin{itemize}
    \item \textbf{Accuracy}~\cite{DBLP:conf/kbse/LiuXYL20, DBLP:conf/iwpc/LinW0MB21}: 
    Between $C_{uc}$ and $C_{gt}$, a score of \textbf{1} is assigned if all tokens match; otherwise, \textbf{0}, 
    following a \textbf{\textit{token-level matching protocol}}:
    (1) Strip formatting (e.g., Markdown markers, language tags, newlines, and prompt phrases);
    (2) Perform camel-case tokenization;
    (3) Convert tokens to lowercase;
    (4) Compare normalized tokens.

    \item \textbf{Aed}~\cite{DBLP:conf/iwpc/LinW0MB21}: 
    Word-level edit distance from $C_{uc}$ to $C_{gt}$: 

    \begin{center}
    $Aed(C_{uc}, C_{gt})$.
    \end{center}

    \item \textbf{Red}~\cite{DBLP:conf/iwpc/LinW0MB21}: 
    Relative edit distance (\textbf{Lower Aed/Red means less editing effort for developers}~\cite{DBLP:conf/kbse/LiuXYL20}): 
    
    \begin{center}
    $Aed(C_{uc}, C_{gt}) / Aed(C_{old}, C_{gt})$.
    \end{center}

    \item \textbf{Bleu-4}~\cite{Papineni02bleu:a}: 
    4-gram similarity between $C_{uc}$ and $C_{gt}$.

    \item \textbf{Meteor}~\cite{banarjee2005}: 
    Word-level alignment metric combining precision, recall, stemming, synonyms, and word order.

    \item \textbf{F1 (Rouge-L)}~\cite{lin2004rouge}: 
    F1 score based on longest common subsequence (LCS) between $C_{uc}$ and $C_{gt}$.

    \item \textbf{SentenceBert}~\cite{reimers2019sentence}: 
    Semantic similarity via cosine distance of embeddings from \texttt{stsb-roberta-large}.
\end{itemize}

\section{
Study Results and Analysis}
\label{sec:expresult}

\begin{table*}[]
\centering
\vspace{-10pt}
\caption{Effectiveness of LLMCup vs. Baselines of Comment Updating}
\label{tab:baselines}
\resizebox{\textwidth}{!}{%
\small
\begin{tabular}{lcccccccc}
\toprule
Method & \#shots & ↑Accuracy & ↓Aed & ↓Red & ↑Bleu-4 & ↑Meteor & ↑F1 & ↑SentenceBert \\
\midrule

CUP & - & 0.177 (116.4\%) & 2.997 (-2.9\%) & 0.938 (-5.7\%) & 0.558 (10.8\%) & 0.788 (4.6\%) & 0.836 (1.9\%) & 0.851 (3.4\%) \\
HebCup & - & 0.257 (49.0\%) & 2.951 (-4.5\%) & 0.858 (-15.5\%) & 0.515 (20.0\%) & 0.788 (4.6\%) & 0.844 (0.9\%) & 0.862 (2.1\%) \\
LLMCup(GPT-4o) & 0,1,3,5 
& \textbf{0.383} 
& 3.084 
& 0.991 
& \textbf{0.618} 
& \textbf{0.824} 
& \textbf{0.852} 
& \textbf{0.880} \\

\bottomrule
\end{tabular}%
}
\small
\raggedright
\textbf{Bold} = best. \textbf{↑}/\textbf{↓} indicate that higher/lower is better. 
\textbf{Positive percentages} indicate performance improvements of LLMCup over the each baseline, while \textbf{negative percentages} denote performance drops.
\textbf{\#Shots}: For baselines, ``-''  denotes no LLM usage. For LLMCup(GPT-4o), a list (0, 1, 3, 5) means the final updated comment is selected among updates using 0-, 1-, 3-, and 5-shot prompts by GPT-4o.

\vspace{-10pt}
\end{table*}

\begin{table*}[]
\centering
\vspace{0pt}
\caption{Effectiveness of LLMCup vs. Basic LLMs}
\label{tab:LLMs Performance}
\resizebox{\textwidth}{!}{
\small
  \begin{tabular}{lcccccccc}
    \toprule
    Method & \#shots & ↑Accuracy & ↓Aed & ↓Red & ↑Bleu-4 & ↑Meteor & ↑F1 & ↑SentenceBert \\
    \hline

    \multirow{4}{*}{Code Llama} & 0 & 0.191 (47.1\%) & 15.329 (71.0\%) & 7.027 (77.5\%) & 0.442 (26.2\%) & 0.723 (7.9\%) & 0.678 (18.4\%) & 0.784 (8.0\%) \\
     & 1 & 0.219 (28.3\%) & 8.364 (46.9\%) & 3.471 (54.5\%) & 0.474 (17.7\%) & 0.706 (10.5\%) & 0.708 (13.4\%) & 0.791 (7.1\%) \\
     & 3 & 0.245 (14.7\%) & 7.266 (38.9\%) & 2.936 (46.3\%) & 0.517 (7.9\%) & 0.751 (3.9\%) & 0.758 (5.9\%) & 0.822 (3.0\%) \\
     & 5 & 0.245 (14.7\%) & 15.163 (70.7\%) & 6.774 (76.7\%) & 0.521 (7.1\%) & 0.757 (3.0\%) & 0.761 (5.5\%) & 0.826 (2.5\%) \\
    LLMCup(CodeLlama) & 0,1,3,5 & \textbf{0.281} & \textbf{4.438} & \textbf{1.578} & \textbf{0.558} & \textbf{0.78} & \textbf{0.803} & \textbf{0.847} \\

    \hline

    \multirow{4}{*}{Deepseek-Coder-V2} & 0 & 0.23 (19.6\%) & 10.036 (56.0\%) & 4.151 (61.7\%) & 0.466 (13.3\%) & 0.723 (5.0\%) & 0.733 (7.9\%) & 0.814 (3.8\%) \\
     & 1 & 0.222 (23.9\%) & 7.301 (39.5\%) & 2.953 (46.1\%) & 0.456 (15.8\%) & 0.699 (8.6\%) & 0.716 (10.5\%) & 0.803 (5.2\%) \\
     & 3 & 0.234 (17.5\%) & 7.441 (40.7\%) & 2.994 (46.9\%) & 0.478 (10.5\%) & 0.726 (4.5\%) & 0.737 (7.3\%) & 0.818 (3.3\%) \\
     & 5 & 0.232 (18.5\%) & 7.086 (37.7\%) & 2.913 (45.4\%) & 0.480 (10.0\%) & 0.728 (4.3\%) & 0.741 (6.7\%) & 0.818 (3.3\%) \\
    LLMCup(Deepseek-Coder-V2) & 0,1,3,5 & \textbf{0.275} & \textbf{4.414} & \textbf{1.591} & \textbf{0.528} & \textbf{0.759} & \textbf{0.791} & \textbf{0.845} \\

    \hline

    \multirow{4}{*}{Gemma} & 0 & 0.139 (28.1\%) & 14.28 (27.9\%) & 6.189 (28.0\%) & 0.378 (13.5\%) & 0.671 (5.5\%) & 0.674 (8.2\%) & 0.783 (3.7\%) \\
     & 1 & 0.108 (64.8\%) & 15.58 (33.9\%) & 6.955 (35.9\%) & 0.349 (22.9\%) & 0.65 (8.9\%) & 0.658 (10.8\%) & 0.775 (4.8\%) \\
     & 3 & 0.109 (63.3\%) & 20.671 (50.2\%) & 9.572 (53.4\%) & 0.334 (28.4\%) & 0.632 (12.0\%) & 0.633 (15.2\%) & 0.764 (6.3\%) \\
     & 5 & 0.098 (81.6\%) & 26.59 (61.3\%) & 12.362 (63.9\%) & 0.311 (37.9\%) & 0.611 (15.9\%) & 0.604 (20.7\%) & 0.752 (8.0\%) \\
    LLMCup(Gemma) & 0,1,3,5 & \textbf{0.178} & \textbf{10.302} & \textbf{4.457} & \textbf{0.429} & \textbf{0.708} & \textbf{0.729} & \textbf{0.812} \\

    \hline

    \multirow{4}{*}{GPT-4o} & 0 & 0.346 (10.7\%) & 4.051 (23.9\%) & 1.43 (30.7\%) & 0.596 (3.7\%) & 0.815 (1.1\%) & 0.832 (2.4\%) & 0.867 (1.5\%) \\
     & 1 & 0.357 (7.3\%) & 4.012 (23.1\%) & 1.386 (28.5\%) & 0.596 (3.7\%) & 0.818 (0.7\%) & 0.833 (2.3\%) & 0.869 (1.3\%) \\
     & 3 & 0.368 (4.1\%) & 3.712 (16.9\%) & 1.279 (22.5\%) & 0.606 (2.0\%) & 0.822 (0.2\%) & 0.84 (1.4\%) & 0.874 (0.7\%) \\
     & 5 & 0.370 (3.5\%) & 3.669 (15.9\%) & 1.260 (21.3\%) & 0.607 (1.8\%) & 0.822 (0.2\%) & 0.841 (1.3\%) & 0.874 (0.7\%) \\
    LLMCup(GPT-4o) & 0,1,3,5 & \textbf{0.383} & \textbf{3.084} & \textbf{0.991} & \textbf{0.618} & \textbf{0.824} & \textbf{0.852} & \textbf{0.88} \\

    \hline

    \multirow{4}{*}{Llama3} & 0 & 0.213 (28.2\%) & 7.22 (40.5\%) & 3.003 (47.8\%) & 0.466 (15.9\%) & 0.728 (5.2\%) & 0.748 (7.6\%) & 0.813 (4.2\%) \\
     & 1 & 0.239 (14.2\%) & 5.757 (25.4\%) & 2.279 (31.2\%) & 0.505 (6.9\%) & 0.743 (3.1\%) & 0.77 (4.5\%) & 0.828 (2.3\%) \\
     & 3 & 0.220 (24.1\%) & 9.057 (52.6\%) & 3.96 (60.4\%) & 0.488 (10.7\%) & 0.742 (3.2\%) & 0.748 (7.6\%) & 0.815 (3.9\%) \\
     & 5 & 0.193 (41.5\%) & 12.425 (65.4\%) & 5.428 (71.1\%) & 0.451 (19.7\%) & 0.718 (6.7\%) & 0.703 (14.5\%) & 0.784 (8.0\%) \\
    LLMCup(Llama3) & 0,1,3,5 & \textbf{0.273} & \textbf{4.295} & \textbf{1.569} & \textbf{0.540} & \textbf{0.766} & \textbf{0.805} & \textbf{0.847} \\

    \hline

    \multirow{4}{*}{Mistral} & 0 & 0.135 (53.3\%) & 15.371 (53.4\%) & 6.760 (57.2\%) & 0.335 (31.9\%) & 0.645 (9.9\%) & 0.630 (15.6\%) & 0.762 (6.6\%) \\
     & 1 & 0.121 (71.1\%) & 11.876 (39.7\%) & 5.331 (45.7\%) & 0.340 (30.0\%) & 0.633 (12.0\%) & 0.634 (14.8\%) & 0.765 (6.1\%) \\
     & 3 & 0.149 (38.9\%) & 10.926 (34.5\%) & 4.772 (39.4\%) & 0.373 (18.5\%) & 0.659 (7.6\%) & 0.659 (10.5\%) & 0.777 (4.5\%) \\
     & 5 & 0.160 (29.4\%) & 10.722 (33.2\%) & 4.723 (38.7\%) & 0.382 (15.7\%) & 0.660 (7.4\%) & 0.661 (10.1\%) & 0.780 (4.1\%) \\
    LLMCup(Mistral) & 0,1,3,5 & \textbf{0.207} & \textbf{7.161} & \textbf{2.893} & \textbf{0.442} & \textbf{0.709} & \textbf{0.728} & \textbf{0.812} \\
    
    \hline

    \multirow{4}{*}{Qwen2.5-Coder} & 0 & 0.249 (23.7\%) & 6.373 (34.2\%) & 2.444 (39.8\%) & 0.515 (9.3\%) & 0.772 (3.1\%) & 0.783 (4.7\%) & 0.841 (2.7\%) \\
     & 1 & 0.271 (13.7\%) & 5.811 (27.8\%) & 2.208 (33.3\%) & 0.527 (6.8\%) & 0.772 (3.1\%) & 0.786 (4.3\%) & 0.844 (2.4\%) \\
     & 3 & 0.280 (10.0\%) & 5.741 (27.0\%) & 2.221 (33.7\%) & 0.532 (5.8\%) & 0.775 (2.7\%) & 0.787 (4.2\%) & 0.845 (2.2\%) \\
     & 5 & 0.285 (8.1\%) & 5.696 (26.4\%) & 2.238 (34.2\%) & 0.535 (5.2\%) & 0.777 (2.4\%) & 0.789 (3.9\%) & 0.848 (1.9\%) \\
    LLMCup(Qwen2.5-Coder) & 0,1,3,5 & \textbf{0.308} & \textbf{4.193} & \textbf{1.472} & \textbf{0.563} & \textbf{0.796} & \textbf{0.820} & \textbf{0.864} \\

    \bottomrule
  \end{tabular}
}
\small
\raggedright
Similar to Table.~\ref{tab:baselines},
\textbf{Positive percentages} indicate performance improvements of LLMCup over the each corresponding basic LLM, while \textbf{negative percentages} denote performance drops.
\textbf{\#Shots}: For basic LLMs, a single integer  indicates the number of retrieved examples. 
\vspace{-10pt}
\end{table*}

\begin{table*}[]
\centering
\vspace{0pt}
\caption{Comparison of LLMCup With Different Rankers}
\label{tab:rankers}
\resizebox{\textwidth}{!}{%
\small
  \begin{tabular}{lccccccccc}
    \toprule
    Method & Ranker & ↑Accuracy & ↓Aed & ↓Red & ↑Bleu-4 & ↑Meteor & ↑F1 & ↑SentenceBert \\
    \hline

    GPT-4o (best @ 5-shot) & - & 0.370 (3.5\%) & 3.669 (15.9\%) & 1.260 (21.3\%) & 0.607 (1.8\%) & 0.822 (0.2\%) & 0.841 (1.3\%) & 0.874 (0.7\%) \\
    \hline
    \multirow{4}{*}{LLMCup(GPT-4o)}
     & Random & 0.360 (6.4\%) & 3.815 (19.2\%) & 1.323 (25.1\%) & 0.601 (2.8\%) & 0.819 (0.6\%) & 0.837 (1.8\%) & 0.871 (1.0\%) \\
     & Self-Rank(GPT-4o) & 0.359 (6.7\%) & 3.856 (20.0\%) & 1.342 (26.2\%) & 0.603 (2.5\%) & 0.820 (0.5\%) & 0.838 (1.7\%) & 0.872 (0.9\%) \\
     & Ranknet & 0.371 (3.2\%) & 3.190 (3.3\%) & 1.020 (2.8\%) & 0.616 (0.3\%) & 0.823 (0.1\%) & 0.850 (0.2\%) & 0.878 (0.2\%) \\
     & CupRank(Ours) & \textbf{0.383} & \textbf{3.084} & \textbf{0.991} & \textbf{0.618} & \textbf{0.824} & \textbf{0.852} & \textbf{0.880} \\
    
    \bottomrule
  \end{tabular}%
}
\small
\raggedright
Similar to Table.~\ref{tab:baselines},
\textbf{Positive percentages} indicate performance improvements of CupRank over the each ranking baseline or GPT-4o (best @ 5-shot), while \textbf{negative percentages} denote performance drops.
\textbf{GPT-4o (best @ 5-shot)}, as a reference, refers to the GPT-4o model that achieves the best accuracy across the 0-, 1-, 3-, and 5-shot settings, specifically under the 5-shot setting.
\vspace{0pt}
\end{table*}

\begin{table*}[]
\centering
\caption{Quantifying the Contribution of Each LLMCup Component}
\label{tab:Ablation}
\resizebox{\textwidth}{!}{%
\small
  \begin{tabular}{lccccccccc}
    \toprule
    Method & \#shots & ↑Accuracy & ↓Aed & ↓Red & ↑Bleu-4 & ↑Meteor & ↑F1 & ↑SentenceBert \\
    \hline

    w/o OpType\&OriginFlag & 0,1,3,5 & 0.375 (-2.1\%) & 3.11 (-0.8\%) & 1.001 (-1.0\%) & 0.617 (-0.2\%) & 0.822 (-0.2\%) & 0.85 (-0.2\%) & 0.879 (-0.1\%) \\

    w/o $\lambda$ of the Loss Function & 0,1,3,5 & 0.374 (-2.3\%) & 3.29 (-6.7\%) & 1.065 (-7.5\%) & 0.617 (-0.2\%) & 0.825 (0.1\%) & 0.851 (-0.1\%) & 0.879 (-0.1\%) \\

    w/o ReprLearningLayer & 0,1,3,5 & 0.369 (-3.7\%) & 3.309 (-7.3\%) & 1.07 (-8.0\%) & 0.612 (-1.0\%) & 0.826 (0.2\%) & 0.848 (-0.5\%) & 0.879 (-0.1\%) \\

    LLMCup(GPT-4o) & 0,1,3,5 & \textbf{0.383} & \textbf{3.084} & \textbf{0.991} & \textbf{0.618} & 0.824 & \textbf{0.852} & \textbf{0.880} \\

    \bottomrule
  \end{tabular}%
}
\small
\raggedright
Similar to Table.~\ref{tab:baselines},
\textbf{Positive percentages} indicate performance improvements of each ablation method over LLMCup(GPT-4o), while \textbf{negative percentages} denote performance drops.
\end{table*}

\subsection{Answer to RQ1: Performance Comparison of LLMCup and State-of-the-Art Methods for Comment Updating}

To compare LLMCup with state-of-the-art baselines (Cup and HebCup), we adopt GPT-4o, top-performing in Table~\ref{Experimental LLMs}, as the default backbone. LLMCup generates four candidate updated comments using 0-, 1-, 3-, and 5-shot prompts and selects the best via ranking.

As shown in Table~\ref{tab:baselines}, LLMCup consistently outperforms CUP and HebCup across most metrics. Accuracy reaches 0.383, improving by 49.0\% over HebCup (0.257) and 116.4\% over CUP (0.177). BLEU-4 improves by 10.8\% and 19.6\%, METEOR by 4.6\%, F1 by up to 1.9\%, and SentenceBert similarity by up to 3.4\%, indicating strong gains in both lexical precision and semantic alignment. 
In contrast, LLMCup shows slightly higher Aed (3.084) and Red (0.991), suggesting increased editing effort—4.5\% and 2.9\% higher Aed than HebCup and CUP, respectively, and Red increases of 15.5\% and 0.4\%. This may reflect LLMCup's tendency to generate more verbose yet semantically rich comments, aligning with improvements in METEOR, F1, and SentenceBert.

\finding{1}{
\textbf{Finding 1.}
Compared to the baselines,
LLMCup increases Accuracy by at least 49.0\%, at most 116.4\%, and 82.95\% on average,
where the average based on the mean of relative improvements over all baselines.
}

\subsection{Answer to RQ2: Evaluating LLMCup Versus Unranked Diverse Prompting with LLMs}

To evaluate the contribution of \textit{CupRank} in LLMCup, we benchmarked it against seven state-of-the-art LLMs (Section~\ref{Experimental LLMs}) using four prompting strategies ( 0-, 1-, 3-, and 5-shot). LLMCup uses CupRank to rank the four candidate comments from these strategies respectively and selects the top-ranked one as the final updated comment.

As shown in Table~\ref{tab:LLMs Performance}, LLMCup consistently outperforms baselines across all LLMs and metrics. For instance, with Code Llama, LLMCup achieves an accuracy of 0.281 compared to the best performance of baselines, a 14.7\% improvement. Aed and Red decrease by 38.9\% and 46.3\%. BLEU-4, Meteor, F1, and SentenceBert similarity improve by 7.1\%, 3.0\%, 5.5\%, and 2.5\%, respectively.
With Deepseek-Coder-V2, at least, accuracy increases from 0.234 to 0.275 (17.5\%), Aed and Red drop by 37.7\% and 45.4\%. BLEU-4, Meteor, F1, and SentenceBert similarity rise by 10.0\%, 4.3\%, 6.7\%, and 3.3\%.
On Gemma, at least, accuracy improves by 28.1\%, Aed and Red reduce by 27.9\% and 28.0\%, while BLEU-4 and F1 increase by 13.5\% and 8.2\%.
For GPT-4o, at least, LLMCup yields accuracy gain of 3.5\%, Aed and Red decrease by 15.9\% and 21.3\%. BLEU-4 improves by 1.8\%, SentenceBert similarity by 0.7\%.

\finding{2}{
\textbf{Finding 2.} CupRank, though trained without GPT-4o's augmented data, consistently improves ranking on GPT-4o's updated comments, highlighting generalization to unseen LLMs in \textit{Data Augmentation}.
}

For Llama3, at least, LLMCup improves accuracy by 14.2\%, reduces Aed and Red by 25.4\% and 31.2\%, and improves BLEU-4, Meteor, and F1 by 6.9\%, 3.1\%, and 4.5\%.
With Mistral, accuracy at least, improves by 29.4\%, Aed by 33.2\%, Red by 38.7\%, BLEU-4 by 15.7\%, and F1 by 10.1\%.
For Qwen2.5-Coder, at least,  accuracy increases by 8.1\%, Aed and Red decrease by 26.4\% and 33.3\%, BLEU-4 improves by 5.2\%, F1 by 3.9\%, and SentenceBert similarity by 1.9\%.

\finding{3}{
\textbf{Finding 3.} Compared to the best performance of basic LLMs, LLMCup improves \textbf{accuracy} by \textbf{3.5\% to 29.4\%}, with the most significant gains on Mistral and Gemma. \textbf{Aed} is reduced by \textbf{15.9\% to 38.9\%}, and \textbf{Red} by \textbf{21.3\% to 46.3\%}. For \textbf{semantic quality}, BLEU-4 improves by \textbf{1.8\% to 15.7\%}, Meteor by \textbf{0.2\% to 7.4\%}, F1 by \textbf{1.3\% to 10.1\%}, and SentenceBert similarity by \textbf{0.7\% to 4.1\%}. These consistent gains demonstrate CupRank’s effectiveness in balancing correctness, minimality, and semantic alignment.
}

\subsection{Answer to RQ3: Comparative Analysis of CupRank and Component Ablation}

This section evaluates the impact of ranking strategies within LLMCup and identifies key contributors to CupRank's success.
Table~\ref{tab:rankers} shows that CupRank consistently outperforms baseline rankers (Random, Self-Rank, RankNet) across all metrics: \textbf{highest accuracy} (0.383, at least 3.2\% improvement), \textbf{lowest Aed} (3.084, at least 3.3\% reduction), and \textbf{lowest Red} (0.991, at least 2.8\% reduction). It also leads in Bleu-4 (0.618, at least 0.3\%), Meteor (0.824, at least 0.1\%), F1 (0.852, at least 0.2\%), and SentenceBert (0.880, at least 0.2\%). Self-Rank (GPT-4o) performs similarly or worse than Random in Accuracy, Aed, and Red, suggesting direct LLM prompting is insufficient. Although RankNet is a strong supervised method, CupRank still outperforms it.

While Random and Self-Rank underperform GPT-4o (best @ 5-shot), both RankNet and CupRank surpass it. Specifically, RankNet improves Accuracy by only 0.3\%, while CupRank achieves a larger improvement of 3.5\%.

\finding{4}{
    \textbf{Finding 4.}
    Random, Self-Rankng, and RankNet are less effective than CupRank in ranking updated comments.
}

To assess component contributions, we performed an ablation study (Table~\ref{tab:Ablation}) on: \textbf{(1) w/o OpType \& OriginFlag}: removing operation type ($a_i$) and origin flag ($b_i$) from $\text{edit token}_i$ (Equation~\ref{eq:edit_token}); \textbf{(2) w/o $\lambda$ of the Loss Function}: neutralizing the $\lambda$ term in the loss (Equations~\ref{eq:logit} and~\ref{eq:loss}); \textbf{(3) w/o Repr Learning Layer}: replacing the representation learning layer (shown in Fig.~\ref{fig:Architecture} and section~\ref{subsec:cuprank}) with a linear layer.

Results show: (1) Removing \textit{OpType \& OriginFlag} causes consistent degradation, where accuracy drops by 2.1\%, Aed and Red increase by 0.8\% and 1.0\%, confirming the need for this data flattening design. (2) Excluding $\lambda$ drops Accuracy by 2.3\% and increases Aed and Red by 6.7\% and 7.5\%, respectively, showing the importance of this loss factor. (3) Removing the \textit{Representation Learning Layer} results in the largest drop: accuracy drops by 3.7\%, Red increases by 8.0\%, and Bleu-4 and F1 drop by 1.0\% and 0.5\%.

\finding{1}{
    \textbf{Finding 5.}
    Ablation results show that input processing (w/o OpType \& OriginFlag), loss design (w/o \(\lambda\)\ of the Loss Function) and learning layers design (w/o Repr Learning Layer) collectively contribute to CupRank’s effectiveness.
}

\subsection{Answer to RQ4: Exploring LLMCup’s Strengths and Weaknesses in Updating Code Comments}
\label{QA4: Types of Update}

To identify when LLMCup performs well or poorly, we examine its accuracy across two dimensions:

\begin{itemize}
\item \textbf{SOURCE Type ($T_s$)}—Whether the comment modification is indicative of code changes: \textbf{Code-Ind} vs. \textbf{Non-Code-Ind}.
\item \textbf{COUNT Type ($T_c$)} — Granularity of changes: \textbf{Single-Token}, \textbf{Single-Sub-Token}, and \textbf{Multi-Tokens}.
\end{itemize}

As shown in Table~\ref{tab:llmcup-types} and Fig.~\ref{fig:type}, LLMCup performs best on \textbf{Code-Ind \& Single-Token} (accuracy 0.828, 2023/2444) and \textbf{Code-Ind \& Single-Sub-Token} (0.778, 1138/1462), indicating strong capability for concise, code-aligned updates.

In contrast, it performs worst on \textbf{Non-Code-Ind \& Multi-Tokens} (0.095, 374/3922), well below the overall average (0.383, 3114/8125), highlighting challenges in understanding or inferring comment changes that lack code
anchoring.

Nonetheless, LLMCup surpasses CUP and HebCup across all update types in absolute correct predictions:
\begin{itemize}
\item 
\textbf{374} on Non-Code-Ind \& Multi-Tokens vs. 74 (improved 5.05 times) of CUP, 9 (improved 41.56 times) of HebCup,
\item 
\textbf{233} on Non-Code-Ind \& Single-Sub-Token vs. 54 (improved 4.31 times) of CUP, 46 (improved 5.07 times) of HebCup,
\item 
\textbf{287} on Non-Code-Ind \& Single-Token vs. 58 (improved 4.95 times) of CUP, 47 (improved 6.11 times) of HebCup,
\item 
\textbf{427} on Code-Ind \& Multi-Tokens vs. 45 (improved 9.49 times) of CUP, 166 (improved 2.57 times) of HebCup,
\item 
\textbf{1138} on Code-Ind \& Single-Sub-Token vs. 703 (improved 1.62 times) of CUP, 1053 (improved 1.08 times) of HebCup,
\item 
\textbf{2023} on Code-Ind \& Single-Token vs. 1254 (improved 1.61 times) of CUP, 1859 (improved 1.09 times) of HebCup.
\end{itemize}

\finding{1}{
\textbf{Finding 6.}
LLMCup excels in \textbf{Code-Ind \& Single-Token} (0.828, 2023/2444) and \textbf{Code-Ind \& Single-Sub-Token} (0.778, 1138/1462), showing strength in code-aligned simple updates.
Its limited area is \textbf{Non-Code-Ind \& Multi-Tokens} (0.095, 374/3922).
However, it consistently outperforms CUP and HebCup across all types by 1.08× to 41.56×.
}

\begin{figure}[h]
     \centering
     \includegraphics[width=0.5\textwidth]{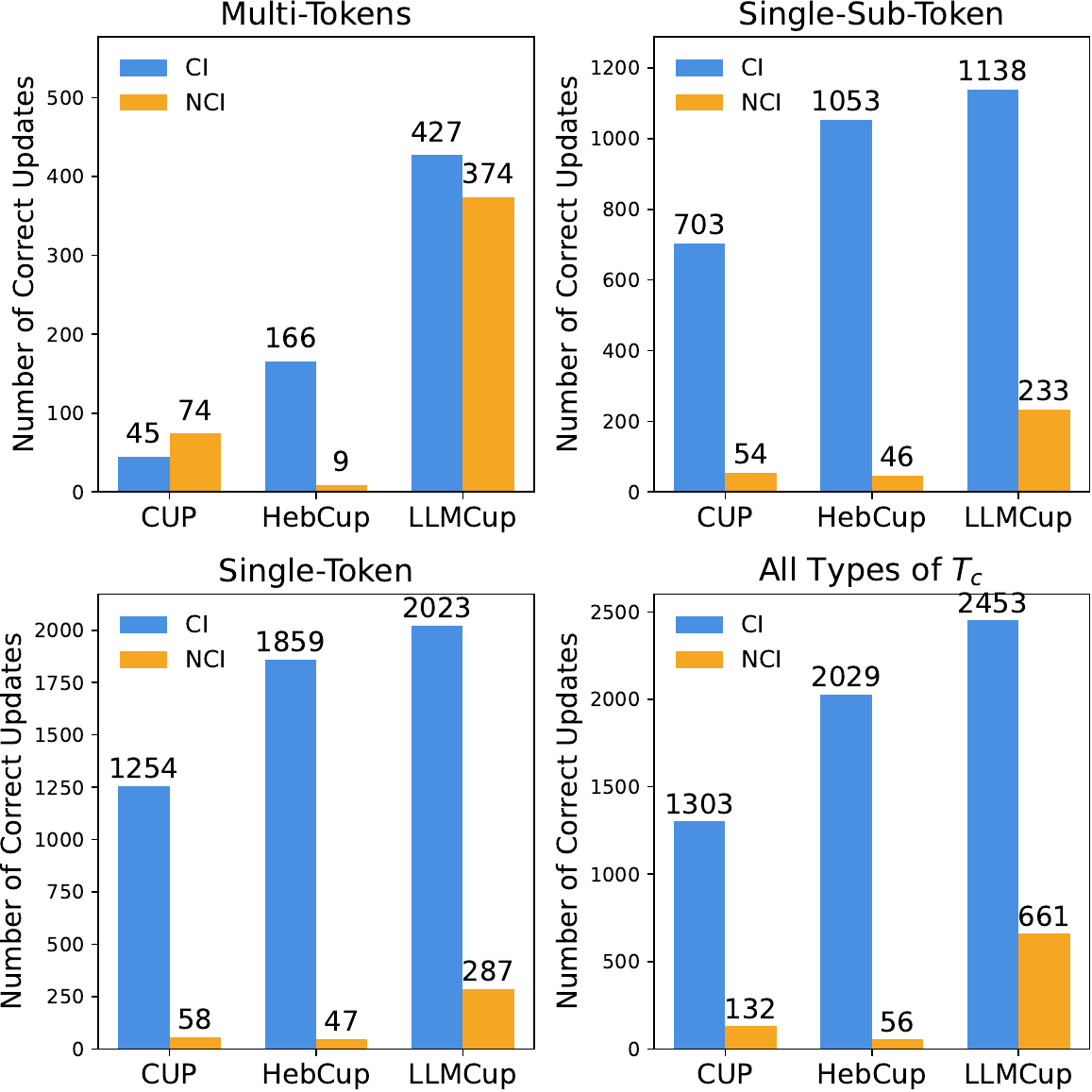}
     \vspace{-10pt}
     \caption{Number of Correct Updates Across Update Types\\
     \footnotesize \textbf{CI} is Code-Indicative while \textbf{NCI} is Non-Code-Indicative.
     }
     \label{fig:type}
     \vspace{-0pt}
\end{figure}

\begin{figure}[]
     \centering
     \vspace{-10pt}
     \includegraphics[width=0.5\textwidth]{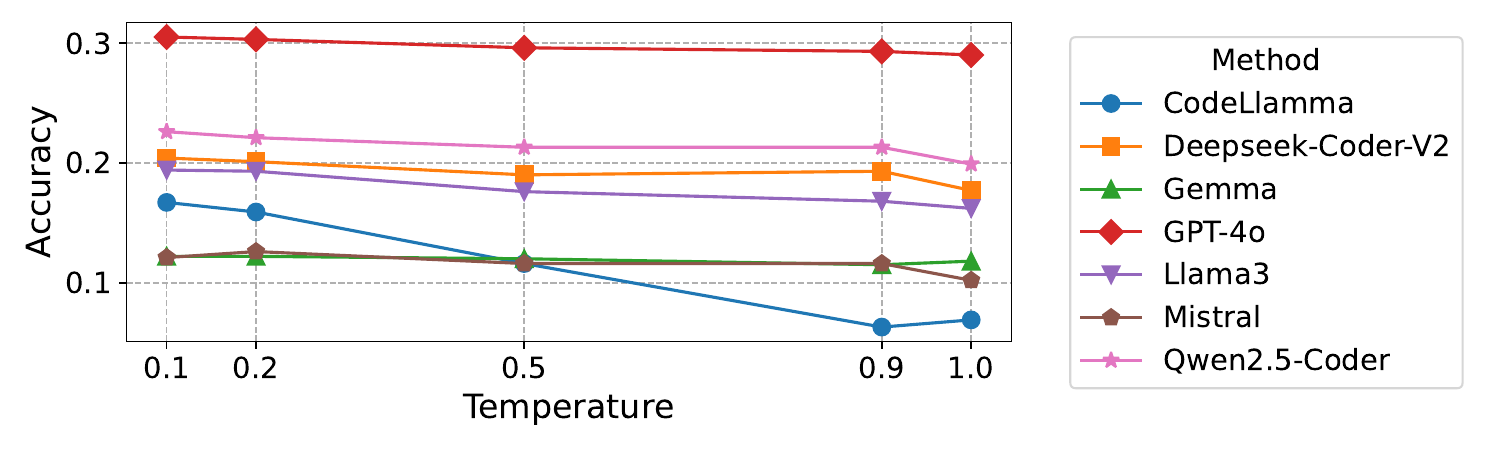}
     \vspace{-20pt}
     \caption{Impact of Temperature Parameter on Comment Update Accuracy for Different LLMs}
     \vspace{-20pt}
     \label{fig:temps}
\end{figure}

\subsection{Answer to RQ5: Impact of Temperature on Comment Updating Performance}  
\label{sec:QA5}

To examine the effect of temperature, we selected five values: 0.1, 0.2, 0.5, 0.9, and 1.0. The lower ones (0.1, 0.2) highlight changes under deterministic settings, while higher ones (0.9, 1.0) illustrate variability at more stochastic levels.

As shown in Fig.~\ref{fig:temps}, accuracy generally declines with increasing temperature. \textbf{GPT-4o} drops from 0.305 at 0.1 to 0.29 at 1.0. \textbf{CodeLlama} shows the steepest decline, from 0.167 to 0.069. \textbf{Llama3} decreases from 0.194 to 0.162.
\textbf{Deepseek-Coder-V2} drops from 0.204 to 0.177, and \textbf{Qwen2.5-Coder} from 0.226 to 0.199, with intermediate values showing steady decrease (e.g., 0.221 at 0.2, 0.213 at 0.5 and 0.9). Both see a net drop of 0.027.
\textbf{Mistral} rises slightly from 0.121 to 0.126 at 0.2, then falls to 0.102 at 1.0 (range: 0.024). \textbf{Gemma} remains stable, ranging from 0.122 to 0.118 (max change: 0.007).

\finding{2}{
\textbf{Finding 7.} \textbf{GPT-4o} achieves the highest accuracy at 0.1 (0.305), followed by \textbf{Qwen2.5-Coder} (0.226) and \textbf{Deepseek-Coder-V2} (0.204). At 1.0, \textbf{CodeLlama} performs worst (0.069). Overall, lower temperatures (0.1--0.2) tend to yield better performance across LLMs.
}

\begin{table}[h]
\caption{LLMCup Accuracy by Update Type}
\centering
\scriptsize
\begin{tabular}{@{}lccc@{}}
\toprule
\diagbox[width=8em]{$T_c$}{$T_s$} & Code-Ind & Non-Code-Ind & Total \\
\midrule

Multi-Tokens & 0.548 (427/779) & \textbf{0.095 (374/3922)} & 0.170 (801/4701) \\

Single-Sub-Token & 0.778 (1138/1462) & 0.284 (233/819) & 0.601 (1371/2281) \\

Single-Token & \textbf{0.828 (2023/2444)} & 0.295 (287/974) & 0.676 (2310/3418) \\

Total & 0.760 (2453/3229) & 0.135 (661/4896) & 0.383 (3114/8125) \\

\bottomrule
\end{tabular}
\label{tab:llmcup-types}
\end{table}

\subsection{RQ6: User Preference for LLMCup vs. Human Comments}

To complement automatic metrics, we conducted a human evaluation to assess whether LLMCup's comments can surpass ground-truth ones on
three  dimension:
(1) \textbf{Consistency}: alignment with the updated code;
(2) \textbf{Naturalness}: language clarity and fluency;
(3) \textbf{Helpfulness}: utility in understanding the method.
We recruited 7 graduate students (average 5 years of Java experience, none involved in this work) to evaluate LLMCup's updated comments. From our dataset, we randomly sampled 100 cases where LLMCup and ground-truth comments differed, ensuring meaningful comparison..

Each instance showed participants an updated method and two anonymized comments, one from LLMCup and one ground truth, in randomized order. Participants rated both on a five-point Likert scale~\cite{1932A}:

As shown in Fig.~\ref{fig:user}, LLMCup comments outperformed human ones on all dimensions:
\textbf{Consistency} (4.14 vs. 3.81), 
\textbf{Naturalness} (3.97 vs. 3.77), 
\textbf{Helpfulness} (4.42 vs. 4.37).

\finding{8}{
\textbf{Finding 8.}
LLMCup's updated comments were on average preferred over ground truth ones across all three dimensions, indicating that LLMCup can outperform humans in comment updating when their outputs differ. This underscores the value of human evaluation as a vital complement to automatic metrics.
}

\begin{figure}[t]
    \centering
    \vspace{-10pt}
    \includegraphics[width=0.35\textwidth]{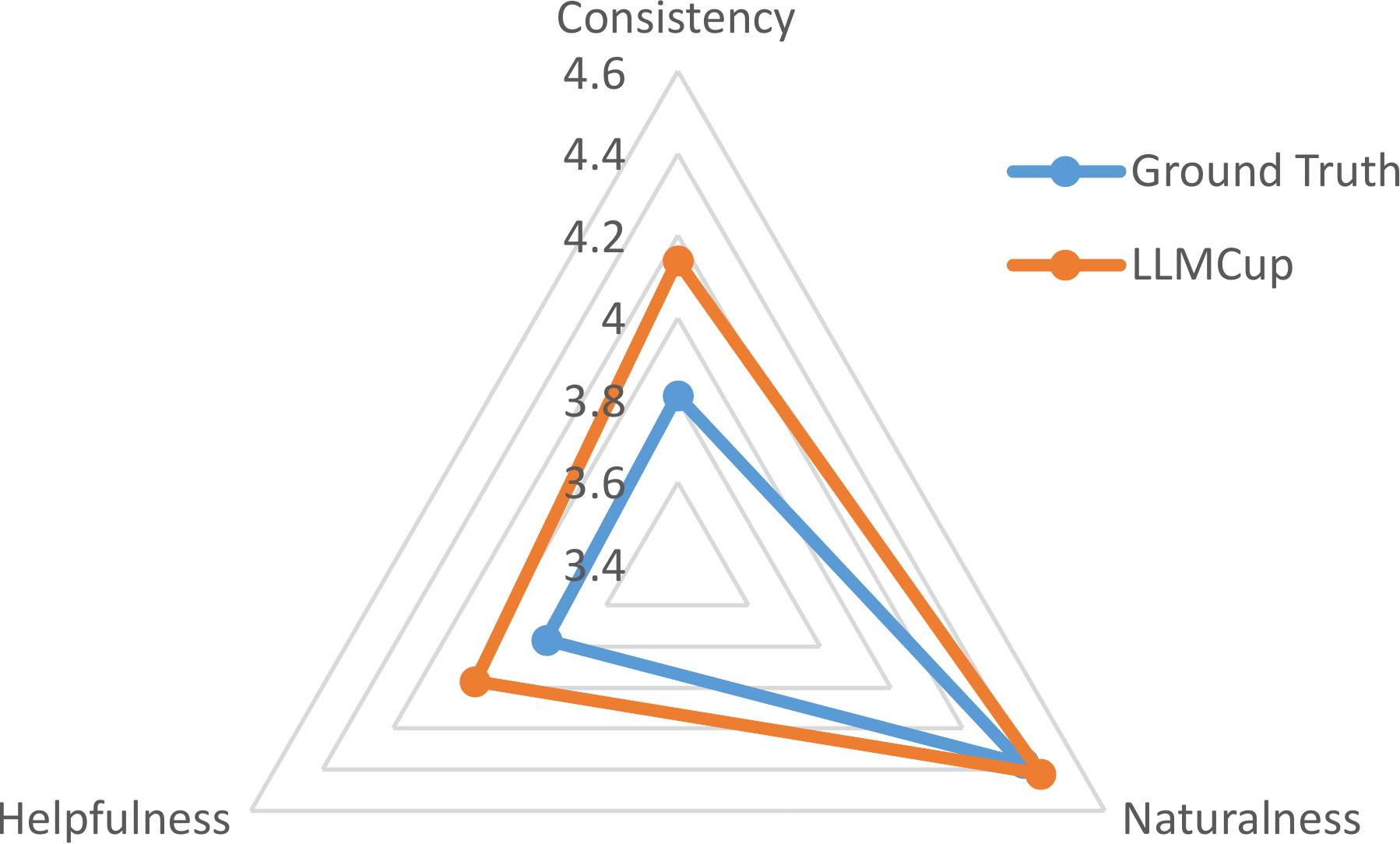}
    \caption{Average user ratings for LLMCup and ground truth comments across three dimensions.}
    \label{fig:user}
    \vspace{-20pt}
\end{figure}

\section{discussion}

\subsection{Limitations and Validity Threats}
\label{sec:threats}

Due to limited resources and budgets, we evaluate LLMCup on seven diverse base LLMs in terms of size, training data, architecture, and accessibility. While representative, this subset may miss some LLM families or top-tier proprietary models, limiting generalizability.
We explore 0-shot, 1-shot, 3-shot, and 5-shot prompts to balance coverage and control. However, not all prompting strategies are considered, posing a validity threat.
Our evaluation uses datasets from open-source Java projects, enabling reproducibility but possibly limiting relevance to proprietary or non-Java systems. Caution is advised when generalizing to such contexts.

\subsection{Practical Implications and Future Work}

LLMCup works with a single LLM using diverse prompts to update and rank comments, suiting contexts where switching LLMs is impractical. Future work will focus on improving ranking methods, especially for non-LLM-generated comments.

\section{Conclusion}
\label{sec:conclusion}

We proposed LLMCup, a novel comment updating approach featuring a new ranking model, CupRank. Experiments show LLMCup outperforms state-of-the-art baselines by 49.0\%–116.9\% in Accuracy, 10.8\%–20\% in BLEU-4, 4.6\% in METEOR, 0.9\%–1.9\% in F1, and 2.1\%–3.4\% in SentenceBert similarity, validating the effectiveness of the LLM-based update-then-rank paradigm. Future work includes ranking updates from diverse LLMs and applying LLMCup in industrial settings.

\bibliographystyle{IEEEtran}

\end{document}